\documentclass[pra,twocolumn, superscriptaddress, footinbib,nobalancelastpage]{revtex4-2}
\usepackage{amsmath}
\usepackage{amssymb}
\usepackage{mathtools}
\usepackage{braket}

\usepackage{graphicx}
\usepackage{calc}
\usepackage{xcolor}

\usepackage[colorlinks,urlcolor=blue,linkcolor=blue,citecolor=blue]{hyperref}
\usepackage{amsmath}
\usepackage{algorithm}
\usepackage{algpseudocode}

\usepackage{mathbbol}
\usepackage{dsfont}
\usepackage{siunitx}
\usepackage{calc}
\usepackage{comment}
\usepackage[normalem]{ulem}

\newcommand{\beq}{\begin{equation}}
\newcommand{\eeq}{\end{equation}}

 \usepackage{ marvosym }
\graphicspath{{figures/}}

\begin{document}

\title{Graph Partitioning into Hamiltonian Subgraphs on a Quantum Annealer}
\author{Eugenio Cocchi}
\affiliation{Quantum Glare Ltd, London, United Kingdom}

\author{Edoardo Tignone}
\affiliation{Quantum Glare Ltd, London, United Kingdom}

\author{Davide Vodola}
\affiliation{Quantum Glare Ltd, London, United Kingdom}
\affiliation{Dipartimento di Fisica e Astronomia dell'Universit\`a di Bologna, I-40127 Bologna, Italy}
\affiliation{INFN, Sezione di Bologna, I-40127 Bologna, Italy}

\begin{abstract}
We demonstrate that a quantum annealer can be used to solve the NP-complete problem of  graph partitioning into  subgraphs containing Hamiltonian cycles of constrained length. We present a method to find a partition of a given directed graph into Hamiltonian subgraphs with three or more vertices,  called vertex 3-cycle cover.  We formulate the problem as a quadratic unconstrained binary optimisation and run it on a D-Wave Advantage quantum annealer.  We test our method on synthetic graphs constructed by adding a number of random edges to a set of disjoint cycles. We show that the probability of solution is independent of the cycle length, and a solution is found for graphs up to 4000 vertices and 5200 edges,  close to the number of physical working qubits available on the quantum annealer.

\end{abstract}

\maketitle

\section{Introduction}
Many combinatorial optimisation problems arising in practical applications are notoriously hard to solve with classical methods \cite{Garey1979}.  Recently, quantum annealers have been considered as potentially faster alternatives for finding solutions to this class of problems in different domains. For example, in logistics they have been employed for job shop scheduling \cite{Venturelli2015}, traffic flow optimisation \cite{Neukart2017, Stollenwerk2017, Inoue2020}, and airport gate assignment \cite{Stollenwerk2018}. In telecommunications quantum annealers have been used for satellite coverage~\cite{Bass2018}, and in chemistry for protein folding \cite{Perdomo-Ortiz2012}. In finance, use cases range from portfolio optimisation \cite{Venturelli2019, Cohen2020} to prediction of financial crashes \cite{Ding2019}.

Finding Hamiltonian cycles, i.e. cycles that visit each vertex exactly once, is another type of combinatorial optimisation problem having several applications, for example in kidney and lung exchange \cite{Constantino2013, Luo2015, Anderson2015}, house allocation~\cite{Abdulkadiroglu1999}, branch selection for cadets \cite{Sonmez2013}, and, more generally, good exchange \cite{Fang2016}. However, so far quantum annealers have not been used to solve these problems.  Here, we consider the partitioning of a directed graph into subgraphs containing Hamiltonian cycles.  More specifically, we focus on the case where the cycle length is required to be  at least three, which makes the problem NP-complete \cite{Garey1979}. This problem is also known as the vertex 3-cycle cover decision problem for directed graphs (3-DCC) \cite{Blaser2001}. 

For solving the problem on the quantum annealer, we cast its cost function and the corresponding constraints as a Quadratic Unconstrained Binary Optimisation (QUBO) problem.  We consider graphs containing disjoint cycles and random edges added and we analyse the probability of finding a vertex 3-cycle cover in a single run on the quantum annealer as a function of the size of the input graph as well as the number of edges added. We find solutions for input graphs up to 4000 vertices and 5200 edges, close to the number of physical working qubits available on the quantum annealer. We find that the dependence of the probability on the system size is stronger when the relative number of random edges added is higher, while it does not depend on the  length of the cycles in the input graph.

This paper is structured as follows. In Sec. \ref{sec_problem} we introduce the relevant definitions and formally define the problem.
In Sec. \ref{sec_qubo_formulation} we show the procedure used to rewrite the problem as a QUBO. In Sec. \ref{sec_implementation_quantum_annealer} we describe the quantum annealing protocol and the procedure used to test the solutions found. In Sec. \ref{sec_results} we present the results obtained with D-Wave Advantage quantum annealer. In Sec. \ref{outlook} we outline possible future developments.

\begin{figure}
\includegraphics[width=0.5\textwidth]{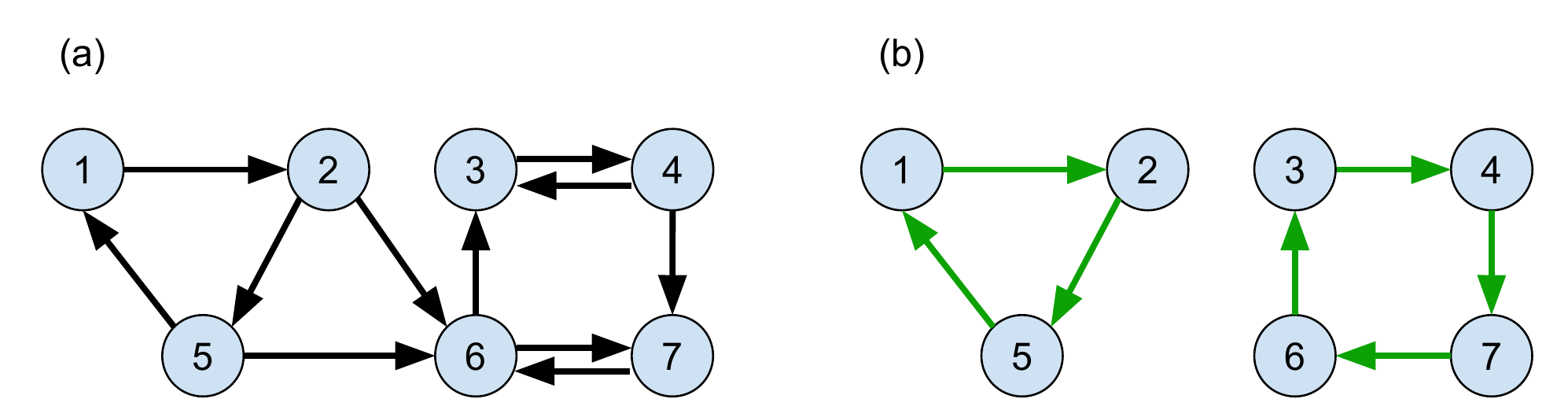}
\caption{\textbf{Partitioning of a directed graph into Hamiltonian subgraphs.} (a) A directed graph is defined by a set of vertices (circles), connected by directed edges (black arrows). (b) A NP-complete version of the problem of partitioning a directed graph  into Hamiltonian subgraphs   consists in choosing the right edges (green arrows) in order to form cycles of length greater than or equal to three, without leaving any vertices unconnected.}
\label{cycles}
\end{figure}

\section{The problem}
\label{sec_problem}
The complexity of the problem of partitioning a graph into Hamiltonian subgraphs strongly depends on the constraints imposed on the size of the subgraphs considered. For example, when no conditions are imposed on the size of the subgraphs, standard matching techniques find a partition in polynomial time~\cite{Edmonds2003}. However, when the subgraphs of the partition are required to have a cardinality greater than or equal to $K$, with $K\geq3$, the problem becomes NP-complete~\cite{Garey1979}. In this paper we consider the $K=3$ problem, that can be formulated as follows: \textit{given a directed graph without self-loops $G = (V,E)$ where $V$ is the set of vertices and $E$ the set of edges, can the vertices be partitioned into disjoint sets $V_1$, $V_2$,  ..., $V_k$ for some $k$ such that each $V_i$ contains at least three vertices and induces a subgraph $G$ that contains a Hamiltonian cycle?}

Figure~\ref{cycles} illustrates the problem for a graph composed of a set of $N_V = 7$ vertices $V = \{1, \dots, 7\}$, and $N_E = 11$ edges $E = \{(1,2), (2,5),(2,6), (3,4), (4,3), (4,7), (5,1), (5,6), (6,3), \\ (6,7), (7,6)\}$  (Fig.~\ref{cycles}a).  
A solution exists for this graph because it does have a partition into Hamiltonian subgraphs with one cycle of length three $V_1 = \{1, 2, 5\}$,  and one cycle of length four $V_2 = \{3, 4, 6, 7\}$ (Fig.~\ref{cycles}b). We stress that, since the problem requires the cycle length to be at least three, the cycles of length two do not appear in the solution.

This problem can be tackled by associating to each edge $(i, j) \in E$ a binary variable $x_{ij}\in\{0,1\}$, that equals 1 when vertices $i$ and $j$ are connected in the solution (see green arrows in Fig.~\ref{cycles}b), 0 otherwise. We note that since the number $N_E$  of existing edges is in general smaller than the number $N_V(N_V-1)$ of possible edges among all the vertices, associating variables $x_{ij}$  only to the existing edges makes the  size of the problem as small as possible.

One can then formulate the problem as:
\begin{equation}
\label{max}
\text{maximise} \sum_{ij} x_{ij} 
\end{equation}
subject to the following constraints: 
\begin{gather}
\label{no_open_cycles}
\sum_{ij} x_{ij} = N_V, \\
\label{max_one_out}
\sum_j x_{ij}  \le 1 \;\;\; \forall \, i  \in V, \\
\label{max_one_in}
\sum_i x_{ij} \le 1 \;\;\; \forall \, j \in V, \\
\label{no_pairs}
x_{ij}+x_{ji}  \le 1   \;\;\; \forall \,  (i, j), (j, i) \in E.
\end{gather}

Constraint \eqref{no_open_cycles} guarantees that the number of edges is equal to the number of vertices. Constraint \eqref{max_one_out} guarantees that for every vertex there is no more than one outgoing  edge, likewise does constraint~\eqref{max_one_in} for the ingoing edges.  Constraints   \eqref{no_open_cycles}-\eqref{max_one_in} guarantee that the solution will be a partition into Hamiltonian subgraphs. Finally, constraint \eqref{no_pairs} ensures that two vertices can be connected to each other by maximum one edge, meaning that cycles of length two are forbidden.

\section{QUBO formulation}
\label{sec_qubo_formulation}
The optimisation problem described by Eqs.~\eqref{max}-\eqref{no_pairs} can be rewritten using the QUBO formalism, which is suitable for a quantum annealer \cite{glover2019tutorial}. In the QUBO formalism we look for a configuration $\mathbf{x}$, where $\mathbf{x}$ is a vector with components    $x_{ij}$, that minimizes the following cost function:
\begin{equation}\label{eqn_qubo_cost}
J(\mathbf{x}) = F(\mathbf{x})  + P(\mathbf{x}).
\end{equation}

The first term of Eq.~\eqref{eqn_qubo_cost} is 
\begin{equation}
\label{eq_free_cost}
F(\mathbf{x}) = - \sum_{ij} x_{ij} 
\end{equation}
and corresponds to Eq.~\eqref{max}.

The second term of Eq.~\eqref{eqn_qubo_cost}  is a penalty term
\begin{equation} 
P(\mathbf{x})  =  P_{\text{out}}(\mathbf{x})  + P_{\text{in}}(\mathbf{x})  + P_{\text{no pairs} }(\mathbf{x}),
\end{equation}
where
\begin{gather}
\label{eqn_qubo_penalties}
P_{\text{out}}(\mathbf{x})  =  \sum_{i}  a_i  \sum_{j, j'>j} x_{ij} x_{ij'}  \\
\label{eqn_qubo_penalties_2}
P_{\text{in}}(\mathbf{x})   = \sum_{j} b_j  \sum_{i, i'>i} x_{ij} x_{i'j}   \\
\label{eqn_qubo_penalties_4}
 P_{\text{no pairs} }(\mathbf{x})    =  c \sum_{i, j>i }   x_{ij} x_{ji} 
\end{gather}

The quantities in Eqs.~\eqref{eqn_qubo_penalties}-\eqref{eqn_qubo_penalties_4} implement the constraints of Eqs.~\eqref{max_one_out}, \eqref{max_one_in} and \eqref{no_pairs}, given that they will be zero when the configuration $\mathbf{x}$ is allowed and positive when the constraint is violated, thus penalising forbidden configurations. 
Equations  \eqref{eqn_qubo_penalties}-\eqref{eqn_qubo_penalties_4} are based on the fact that for any two binary variables $y$ and $z$,  the constraint $y+z\leq1$ is equivalent to $y \cdot z =0$. We note that it is not necessary to encode constraint  \eqref{no_open_cycles}  as a penalty term, since it has the same functional form as Eq.~\eqref{max}, and checking the solution found will be sufficient.

When translating the constraints into penalties one needs to choose the penalty constants $a_i$, $b_j$, $c$ large enough compared to the strength of the term $\sum_{ij} x_{ij}$. However, due to the hardware implementation, these cannot be chosen  arbitrarily large. An optimal choice for the penalty constants is presented in Appendix \ref{appendix_constraints}.

\section{Implementation on a quantum annealer}
\label{sec_implementation_quantum_annealer}

 \subsection{Quantum annealing}
The constructed QUBO problem  is solved on a D-Wave Advantage quantum annealer, containing 5436 physical working qubits.  The starting point for the quantum routine used is a quantum state that corresponds to the ground state of a drive Hamiltonian 
\begin{equation} 
H_0 = -\sum_{\ell} \sigma^x_\ell.
\end{equation}
This Hamiltonian is slowly changed to the problem Hamiltonian $H_1$ whose ground state represents the state with lowest energy for the QUBO problem.  Its general form is that of an Ising-like Hamiltonian: 
\begin{equation}
\label{H1}
H_1 = \sum_{\ell} h_\ell \sigma_\ell^z + \sum_{\ell \ell'} t_{\ell \ell'} \sigma_\ell^z \sigma_\ell^z\ .
\end{equation}
In $H_0$ and $H_1$, $\ell$ and $\ell'$ are indices denoting the position of the physical qubits in the hardware, $\sigma^{x,z}_\ell$  are Pauli operators, and the parameters $h_\ell$ and $t_{\ell\ell'}$ are set by the QUBO problem.

The total evolution can then be modelled via the Hamiltonian $H(s) = A(s) H_0 + B(s) H_1$, where $A(s)$ and $B(s)$ are a decreasing and increasing function of the dimensionless schedule parameter $s\in [0, 1]$, respectively. The $A(s)$ and $B(s)$ functions are fixed by the hardware, while the time variation of $s$ is controlled by the programmed schedule. 
In our experiment we use a schedule having a total duration of  300 $\mu$s, composed of an initial annealing where the parameter $s$ is linearly ramped up from 0 to 0.4 in 80 $\mu$s, followed by a pause at $s=0.4$ lasting for 100 $\mu$s, and a second part of the annealing consisting of a linear ramp of $s$ from 0.4 to 1 in 120 $\mu$s.

The mapping of the QUBO problem in Eq.~\eqref{eqn_qubo_cost} to the Ising-like Hamiltonian in Eq. \eqref{H1} is done by identifying the two states $\{0,1\}$ of the variables $x_{ij}$  with the two eigenstates of the $\sigma_\ell^z$ operator of the qubit $\ell$.  Since the physical qubits in the quantum processor are not fully connected to each other,  each logical qubit  is embedded into a chain of one or more physical qubits. To find such an embedding, we use the \textit{minorminer} algorithm~\cite{Cai2014} provided by D-Wave, with its default parameters.

\subsection{Protocol for solving the partitioning problem}
\label{protocol_paritioning}
\begin{figure}
\includegraphics[width=0.5\textwidth]{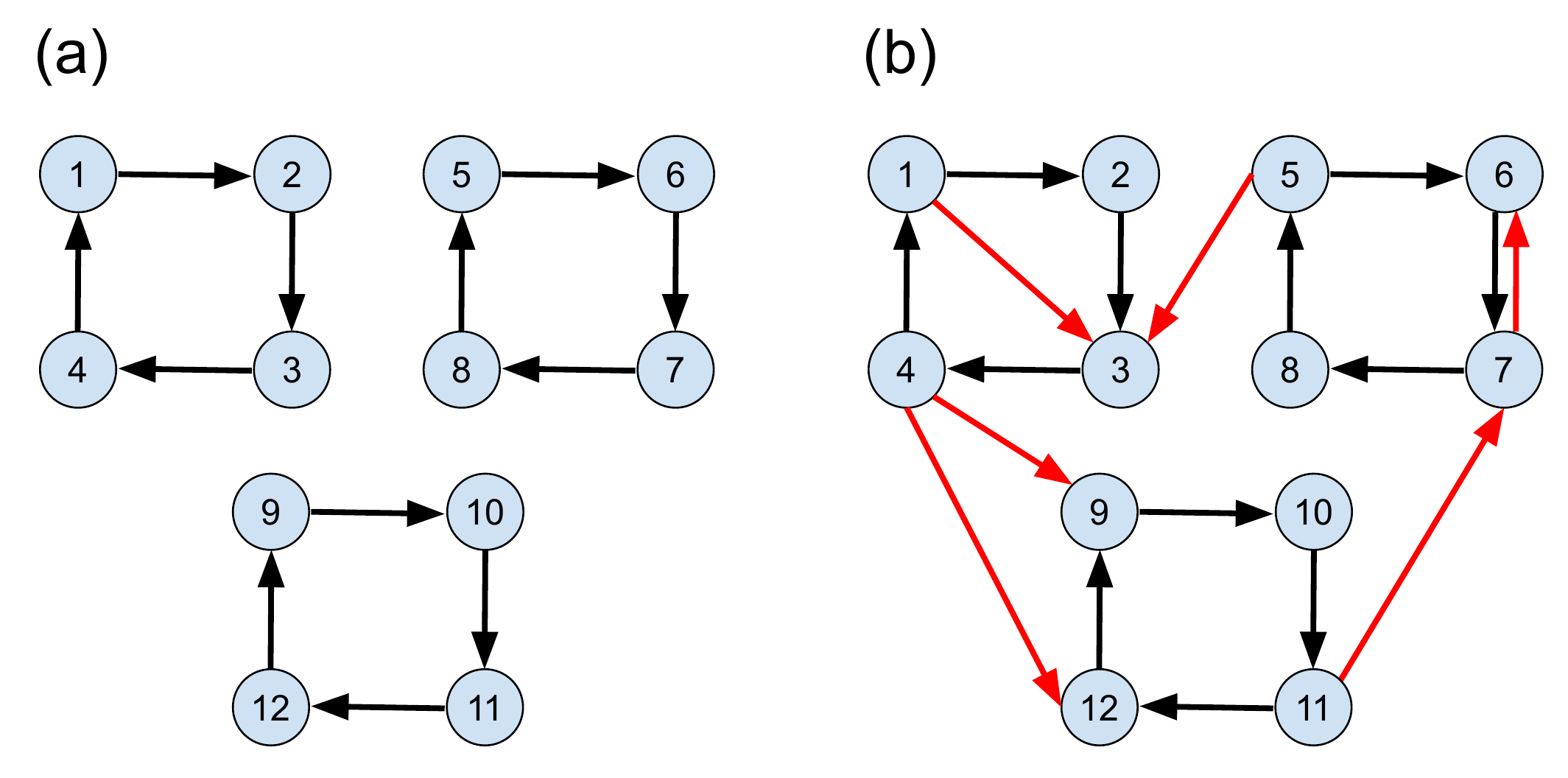}
\caption{\textbf{Construction of the input graphs for the partitioning problem.} (a) An example of graph $G_0$ with parameters $n=3$,  $L=4$, and an equal number of vertices and edges $N_{V} = N_{E}=12$. (b) A graph $G$ generated from $G_0$, by adding $N_\text{noise}=6$ new edges (red arrows). For $G$ we have $N_{V}=12$, $N_{E}=18$.}
\label{add_noise}
\end{figure}

In this section we present the steps for the protocol we use to solve the problem on the quantum annealer: (i) graph construction, (ii) problem submission, (iii) check of the output.

\begin{algorithm}[H]
  \caption{Solution check. The pseudocode describes the function used to verify whether a graph $G'= (V, E')$, with $E' \subseteq E$, is a partition of a graph $G= (V, E)$ into Hamiltonian subgraphs with three or more vertices.}
  \label{algo_sol_check}
  \textbf{Input:} Graphs $G'= (V, E')$ and  $G= (V, E)$\\
  \textbf{Output:} True if $G'$ is a partition of $G$ into Hamiltonian subgraphs with three or more vertices, False otherwise\\
  \textbf{Procedure:}
   \begin{algorithmic}[1]
   \Repeat
    \State Initialise two empty sets $C_{V} = \{\}$ and $C_{E} = \{\}$
   \State Pick a vertex $v_\text{start}  \in V$
   \State Assign $v_\text{from}:=v_\text{start}$
   \Repeat
   \State Find the edges $E_\text{from} \subseteq E'$ going out of $v_\text{from}$
   \If{$E_\text{from}$ has exactly one element} 
      \State Add the vertex $v_\text{from}$ to $C_{V}$
      \State Add the edge $e$ in $E_\text{from} $ to $C_{E} $
   \Else   
  \State \Return False
  \EndIf 
   \State Assign $v_\text{to}$ to the vertex pointed at by $e$
              \If{($v_\text{to}\neq v_\text{start}$) and ($v_\text{to}\in C_V$ or $C_{E} = E'$)}
  \State \Return False
  \EndIf
   \State Assign $v_\text{from}:=v_\text{to}$
   \Until {$v_\text{to}=v_\text{start}$}
   \If{$C_{V}$ has exactly 2 elements} 
   \State \Return False
   \ElsIf{$C_{V}$ has 3 elements or more} 
    \State Remove from $V$ the vertices contained in $C_V$
    \State Remove from $E'$ the edges contained in $C_E$
   \EndIf
      \Until {no vertices are left in $V$ and no edges are left in $E'$}
        \State \Return True
    \end{algorithmic}
   \end{algorithm}

(i) We start by generating a graph $G_0$ composed of $n$ disjoint cycles of length $L$ (Fig.~\ref{add_noise}a). This contains $nL$ vertices and $nL$ edges.
A new graph $G$ is generated starting from $G_0$, by introducing noise, i.e. adding $N_\text{noise}$ new randomly chosen edges that connect the existing vertices (Fig.~\ref{add_noise}b). The so-constructed graph $G$ will always admit $G_0$ as solution, even though additional solutions might also appear when the amount of noise is large.

(ii) The graph $G$ is then transformed into a QUBO problem as explained in the previous sections and submitted to the quantum annealer.  The annealing schedule is run 100 times, and the frequency of the  final states obtained is computed.

(iii) Using Algorithm \ref{algo_sol_check} we check if the lowest energy state (or states in the degenerate case) corresponds to a partition of $G$ into Hamiltonian subgraphs containing cycles of length three or more. If that is the case, the probability $P_\text{sol}$ of finding a solution is equal to frequency of the lowest energy state (or the sum of the frequencies of the lowest energy states in the degenerate case), if not $P_\text{sol}=0$. We note that the solution check algorithm runs in polynomial time. The only step that is proportional to the size of the problem is line~6 in Algorithm \ref{algo_sol_check}, which is a $\mathcal{O}(N_E)$. That step  is executed at maximum $N_V$ times, which makes the overall algorithm a $\mathcal{O}(N_EN_V)$.

To collect statistics on $P_\text{sol}$, we repeat the steps (i) to~(iii) 50 times and average $P_\text{sol}$ over the 50 repetitions to obtain $\bar{P}_\text{sol}$, that represents the average probability to find a solution with a single run (i.e. a single annealing schedule) on the quantum annealer.

\section{Results}
\label{sec_results}
\begin{figure}
\includegraphics[width=0.5\textwidth]{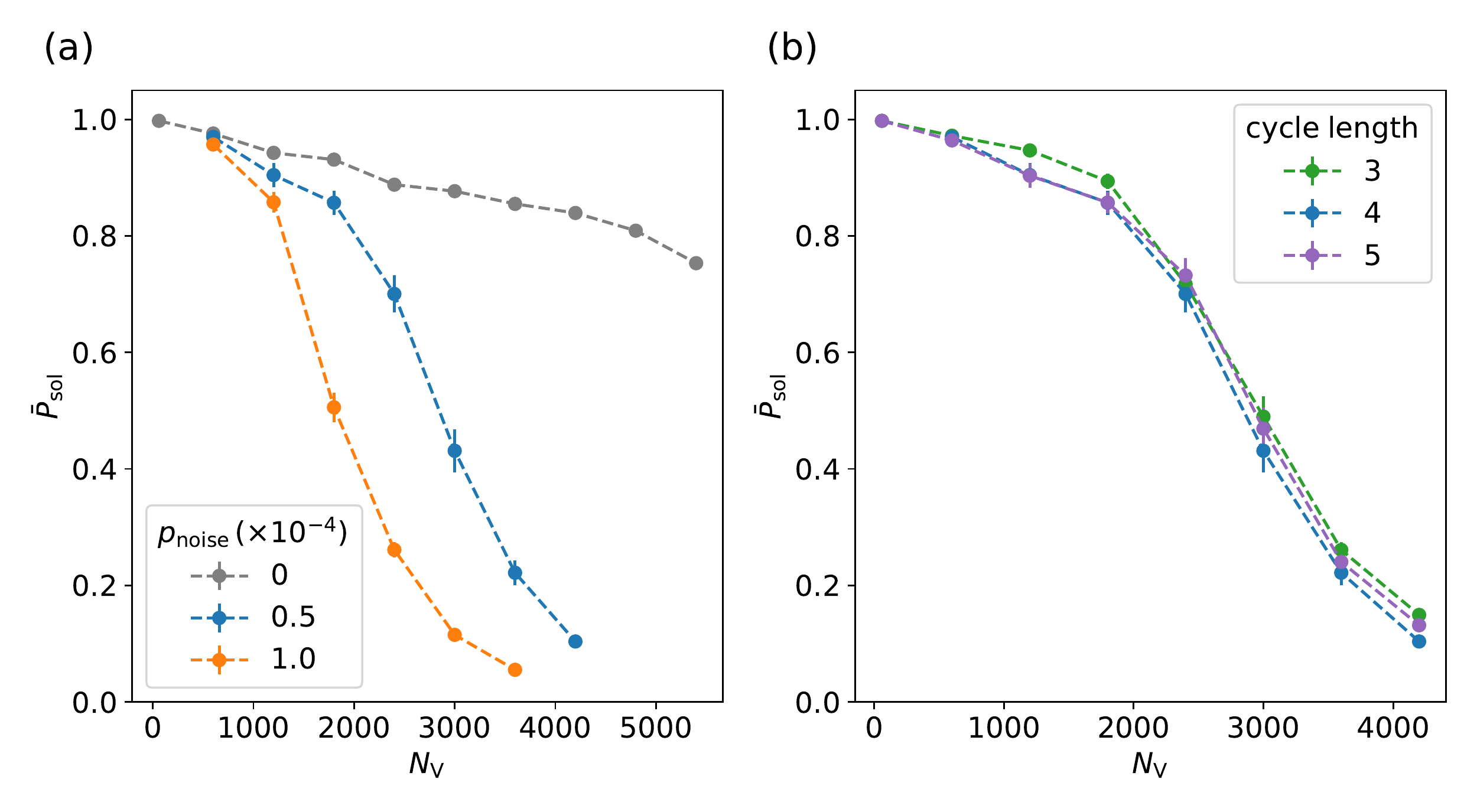}
\caption{\textbf{Single-run solution probability as a function of the number of vertices.} (a) $\bar{P}_\text{sol}$ is shown as a function of $N_V$ for cycles of length 4 for different $p_\text{noise}$. 
For all curves the probability to find a solution in a single run on the quantum annealer decreases as the system size increases.
With noise, the dependence on the system size is stronger: for $N_V=4200$  and $p_\text{noise}=0.5\times 10 ^{-4}$ (i.e. $N_\text{noise}=882$)  the probability is  $\bar{P}_\text{sol} = 0.104(9)$, and for $N_V=3600$  and $p_\text{noise}=1\times 10 ^{-4}$ (i.e.  $N_\text{noise}=1295$), the  probability is $\bar{P}_\text{sol} = 0.055(8)$. (b)  $\bar{P}_\text{sol}$ is shown as a function of $N_V$ for $p_\text{noise}=0.5\times 10 ^{-4}$  for different cycle lengths. The probability exhibits the same behaviour as a function of the problem size regardless of the cycle length used in the input graph configuration. In both panels, the error bars show the standard error on the mean.}
\label{varying_n_vertices}
\end{figure}

We start by fixing the fraction $p_\text{noise}$ of the maximum allowed number of additional edges for the given number of vertices $N_{V}$, and set $N_\text{noise} = \text{round}(p_\text{noise} \  N_{V} (N_{V}-2))$.
In the simplest scenario, i.e. $p_\text{noise}$ = 0, the quantum annealer easily finds a solution regardless of the problem size (grey points in Fig.~\ref{varying_n_vertices}a). Even when considering a problem with a very large number of vertices and edges, $N_V = N_E = 5400$, where the graph size is very close to the total number of working physical qubits available (5436),  the probability to find a solution with a single run on the quantum annealer is  $\bar{P}_\text{sol}=75(1)\%$.  We observe that $\bar{P}_\text{sol}$ decreases with $N_V$ with a slope that strongly depends on $p_\text{noise}$, and the quantum annealer finds a solution up to  $N_V=4200$ for $p_\text{noise}=0.5\times 10 ^{-4}$  and up to $N_V=3600$  for $p_\text{noise}=1\times 10 ^{-4}$ (blue and orange points in Fig.~\ref{varying_n_vertices}a). 

Figure \ref{varying_n_vertices}b shows that the single-run solution probability $\bar{P}_\text{sol}$ does not depend on the cycle length. This can be explained by the fact that the dimension of the combinatorial space depends on the number of all the possible paths in the input graph, which is determined only by $ N_{V}$  and $p_\text{noise}$.

\begin{figure}
\includegraphics[width=0.5\textwidth]{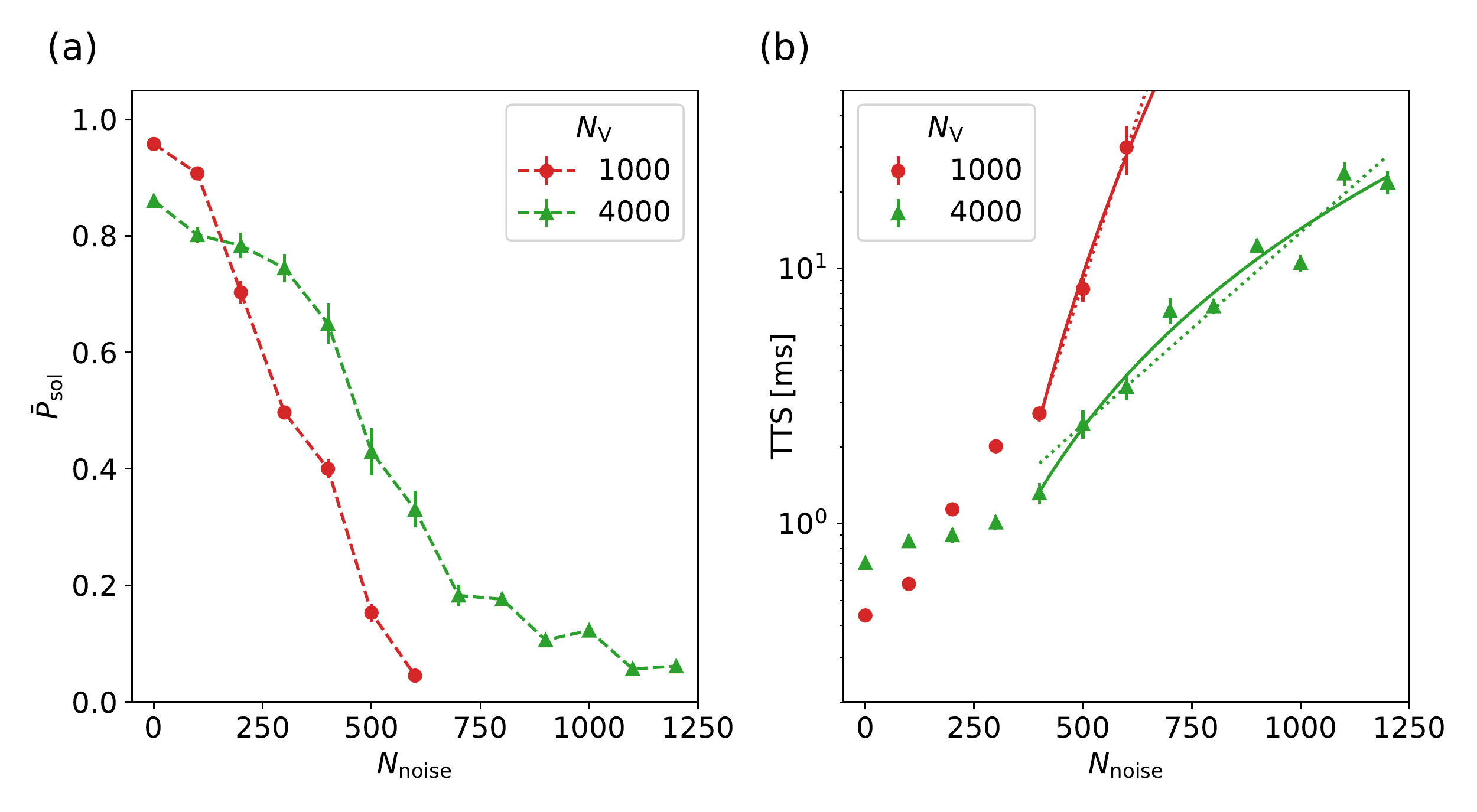}
\caption{\textbf{Single-run solution probability and time to solution as a function of the number of added random edges.} (a) $\bar{P}_\text{sol}$ is shown as a function of $N_\text{noise}$ for cycles of length 4 and different $N_\text{V}$. The single-run solution probability decreases as the number of random edges becomes larger. For larger sizes of the problem,
$\bar{P}_\text{sol}$ goes to zero more slowly. (b) TTS shown for the same points as panel~(a). The dotted lines show exponential fits to the data in the range $N_\text{noise}\geq400$  with equation $\log(\text{TTS}/1\text{ms}) = a + bN_\text{noise}$ and the solid lines show power law fits with equation $\log(\text{TTS}/1\text{ms}) = a + b \log(N_\text{noise})$. 
}
\label{varying_n_noise}
\end{figure}

    To further investigate the dependence on the combinatorial complexity, we now fix the number of vertices $N_V$ and vary the number of added edges $N_\text{noise}$. To give an intuition on how the complexity of the problem scales with $N_\text{noise}$, we can think of a classical approach based on  Algorithm \ref{algo_sol_check}. Without any edges added, there is only one possible simple path. When we add one edge of noise, there will be one vertex with two outgoing edges: this bifurcation gives rise to two different simple paths. Each of them could be then fed into Algorithm   \ref{algo_sol_check} to check whether it is a solution. However, when the number of added edges increases, the number of simple paths to be checked can be up to $2^{N_\text{noise}}$,  and therefore finding a solution with this classical procedure becomes exponentially hard.
    
    In Fig.~\ref{varying_n_noise} we show the results obtained with the quantum annealer. For a small amount of added edges ($N_\text{noise} \lesssim 100$) the single-run solution probability is higher for the smaller system size explored (Fig.~\ref{varying_n_noise}a).  However, as $N_\text{noise}$ is increased, for a given number of added edges it becomes easier to find a solution for a system where the size is larger. We note that, regardless of the value of the single-run solution probability, the same problem can be submitted multiple times in order to find a solution at least once with arbitrarily high probability.  Fixing the desired probability to 99\%, the time needed to find a solution is given by
 \cite{Albash2018} 
\begin{equation}
\text{TTS}  = (t_\text{anneal} + t_\text{pause}) \frac{\log(1-0.99)}{\log(1-\bar{P}_\text{sol})},
\end{equation} 
where the first term is the single-run time (i.e. the sum of the times used in the schedule of the quantum annealer $t_\text{anneal} = 200 \ \mu \text{s}$ and $t_\text{pause} = 100 \ \mu \text{s}$) and the second term is the number of necessary runs to find a solution with the desired probability.

In Fig.~\ref{varying_n_noise}b we show the behaviour of the time to solution  $\text{TTS} $ as a function of the number of added random edges. We fit the large $N_\text{noise}$ behaviour of TTS with an exponential and power law functions. The fit parameters are reported in Table \ref{fit_parameters}. It is clear from the plots that for the range explored TTS is compatible with either fit.  However, even in the exponential case, the scaling is much slower than the one of the classical procedure explained earlier, that scales as  $\exp{(N_\text{noise}\log 2)}$.

\begin{table}
\begin{tabular}{c|c|c|c}
 & $a$ & $b$ \\
\hline
power law fit $N_V = 1000$ &$-34(3)$&5.9(6)\\
power law fit $N_V = 4000$ &$-15(1)$&2.6(2)\\
\hline
exponential fit $N_V = 1000$ &$-3.8(2)$&0.0120(4)\\
exponential fit $N_V = 4000$ &$-0.8(3)$&0.0035(3)\\
\end{tabular}

\caption{\textbf{Fit parameters}. The table reports the parameters obtained  from the fits to the data shown in Fig.  \ref{varying_n_noise}b. For the exponential fit, the parameter $b$  is much less than $\log2$, expected for the $2^{N_\text{noise}}$ classical scaling described in section~\ref{sec_results}.
}
\label{fit_parameters}
\end{table}

\section{Outlook}
\label{outlook}
Possible extensions of the problem presented here can be considered for graphs where weights are assigned to the edges, or where different constraints on the cycle length are present.

We point out that, if self-loops are included in the construction of the problem and the constraint on the cycle length is lifted, our method could  be used to compute the permanent of a matrix, since finding all the cycle covers of a graph is equivalent to computing the permanent of its adjacency matrix  \cite{Rudolph2009}.

\vspace{0.5cm}
\noindent \textbf{Additional information:}
The code used to generate the results presented in this paper is available at \url{https://github.com/quantumglare/quantum_cycle}.
Further information can be requested at \href{mailto:info@quantumglare.com}{\texttt{info@quantumglare.com}}.

\appendix
\section{Constraints}
\label{appendix_constraints}
In order to properly set the penalty constants in Eqs.~\eqref{eqn_qubo_penalties}-\eqref{eqn_qubo_penalties_4} we proceed with an analysis of the different cost terms in Eq.~\eqref{eqn_qubo_cost}, that allows us to choose the penalty constants as small as possible.

We require  the cost $J(\mathbf{x}_a)$ of an allowed configuration~$\mathbf{x}_a$ that satisfies all constraints  to be lower than the cost $J(\mathbf{x})$ of any configuration $\mathbf{x}$ that violates at least one constraint, i.e.
\begin{equation}
\label{eq_constraint_cost_inequality}
J(\mathbf{x}) > J(\mathbf{x}_a) \quad \forall \  \mathbf{x}, \mathbf{x}_a.
\end{equation}
\vspace{0.1cm}

Let us first consider the constraint on the number of outgoing edges given in Eq.~\eqref{max_one_out}, whose corresponding penalty is given in Eq.~\eqref{eqn_qubo_penalties}. Any configuration that violates only that constraint  can be  decomposed as $\mathbf{x} = \mathbf{x}_a + \mathbf{x'}$, where $\mathbf{x'}$ is a vector whose only elements equal to~1 are those corresponding to the additional edges. From equation \eqref{eq_constraint_cost_inequality} it follows that, for every vertex $i$,
\begin{equation}
\label{eq_constraint_cost_inequality_max_one_out}
 a_i  \sum_{j, j'>j} x_{ij} x_{ij'}  > \sum_{j} x'_{ij}.
\end{equation}
Equation \eqref{eq_constraint_cost_inequality_max_one_out} is satisfied by setting:
\begin{equation}
\label{eq_constraint_cost_inequality_max_one_out_epsilon_raw}
a_i = \max_{\mathbf{x}, \mathbf{x}'} \left( \frac{\sum_{j} x'_{ij}}{ \sum_{j, j'>j} x_{ij} x_{ij'} }  \right) + \epsilon,
\end{equation}
where $\epsilon$ is an arbitrarily small positive constant, which makes $a_i$ an optimal choice. The number $N_{\mathbf{x'},i}$ of non-zero elements in $\mathbf{x}'$ varies from $1$ to  the total number ${N_{\text{out},\,i}}$ of outgoing edges from vertex $i$ in the original graph. In terms of $N_{\mathbf{x'},i}$ Eq.~\eqref{eq_constraint_cost_inequality_max_one_out_epsilon_raw} becomes
 \begin{equation}
 \label{eq_a_1_appendix}
a_i = \max_{N_{\mathbf{x'},i}} \left( \frac{N_{\mathbf{x'},i}}{\binom{N_{\mathbf{x'},i}+1}{2}}  \right) + \epsilon
\end{equation}
where in the denominator the round brackets denote the binomial coefficient.
The maximum is achieved for $N_{\mathbf{x'},i}=1$, giving $a_i=1+\epsilon$ for all vertices $i$ of the graph that might violate the constraint. For all other vertices we simply set it to zero, leading to
 \begin{equation}
a_i = \begin{cases}  
1 + \epsilon  & \text{if} \ N_{\text{out},\,i} > 1,  \\
0 & \text{otherwise}.
\end{cases}
\end{equation}

Likewise, for the constraint on the number of ingoing edges in Eq.~\eqref{max_one_in}, which corresponds to the penalty in Eq.~\eqref{eqn_qubo_penalties_2}, we set
 \begin{equation}
b_i = \begin{cases}  
1 + \epsilon & \text{if} \ N_{\text{in},\,i}  > 1, \\
0 & \text{otherwise},
\end{cases}
\end{equation}
where  ${N_{\text{in},\,i}}$ is the number of ingoing edges to vertex $i$ in the original graph.

A similar reasoning for the constraint to forbid pairs in Eq.~\eqref{no_pairs}, which corresponds to the penalty in Eq.~\eqref{eqn_qubo_penalties_4}, leads to
 \begin{equation}
c = 2 + \epsilon.
\end{equation}

\end{document}